\ifx\pdfoutput\undefined
\documentstyle[prl,aps,amsmath,amssymb,multicol,epsfig]{revtex}
\else
\documentstyle[prl,aps,amsmath,amssymb,multicol,epsfig]{revtex}
\fi
\DeclareMathOperator{\EllipticK}{K}
\DeclareMathOperator{\EllipticE}{E}
\DeclareMathOperator{\JacobiSN}{sn}
\DeclareMathOperator{\JacobiDN}{dn}
\newcommand{\gtsd}{{\mathrm d}}
\newcommand{\Jacobim}{{\mathit m}}
\newcommand{\gtsu}{u}
\newcommand{\gtsSign}{\varepsilon}
\newcommand{\gtsConst}{{\mathsf c}}
\newcommand{\gtsCst}{c}
\newcommand{\gtsSurface}{{\mathcal S}}
\newcommand{\gtsOR}{\overline{\mathit R}}
\newcommand{\gtsIR}{\underline{\mathit R}}
\newcommand{\gtsHm}{{\mathcal H}}
\newcommand{\gtsJ}{{\mathit J}}
\newcommand{\gtsKc}{{\mathit k}_{c}}
\newcommand{\gtsJr}{\kappa}
\newcommand{\gtsMTheta}{\Theta}
\newcommand{\gtsMPhi}{\Phi}
\newcommand{\gtsq}{q}
\newcommand{\gtsQ}{Q}
\newcommand{\gtsm}{\widetilde{\Jacobim}}
\newcommand{\gtsSG}{\alpha}
\newcommand{\gtsK}{K}
\newcommand{\gtsE}{{\mathrm E}}
\newcommand{\gtsTE}{\underline{\mathrm E}}
\newcommand{\gtsEr}{{\mathcal E}}
\newcommand{\gtsTL}{\Lambda}
\newcommand{\gtsMC}{{\mathit H}}
\newcommand{\gtsLm}{\Jacobim}
\newcommand{\gtsLA}{A}
\newcommand{\gtsLB}{B}
\newcommand{\gtsLj}{j}
\newcommand{\gtsIL}{{\mathfrak L}}
\newcommand{\gtsrd}{\lambda}
\begin{document}
\title{Heisenberg Spins on an Elastic Torus Section%
  \thanks{\textsc{Physics Letters A} \textbf{248}, 439 (1998) %
    [\texttt{cond-mat}/9809266]}}
\author{J.~Benoit\thanks{E-mail: benoit@u-cergy.fr}%
  and R.~Dandoloff\thanks{E-mail: rossen@u-cergy.fr}}
\address{%
  Laboratoire de Physique Th\'eorique et Mod\'elisation,
  Universit\'e de Cergy-Pontoise,\\
  2, avenue A.~Chauvin,
  95302~Cergy-Pontoise,France}
\date{Received 5 May 1998; %
  revised manuscript received 13 August 1998; %
  accepted for publication 14 August 1998\\
  Communicated by A.~R. Bishop}
\maketitle
\begin{abstract}
Classical Heisenberg spins in the continuum limit
(i.e. the nonlinear $\sigma$-model) are studied on an
elastic torus section with homogeneous boundary conditions.
The corresponding rigid model exhibits topological soliton
configurations with geometrical frustration
due to the torus eccentricity.
Assuming small and smooth deformations allows to find shapes of
the elastic support by relaxing the rigidity constraint:
an inhomogeneous Lam\'e equation arises.
Finally, this leads to a novel geometric effect:
a \textit{global shrinking} with swellings.
\copyright 1998 Elsevier Science B.~V.
\end{abstract}

\begin{multicols}{2}
Curved magnetic structures abound in nature:
either made out of magnetic materials
or enclosing magnetorheological fluids.
To explore their magnetoelastic properties, it is convenient
to treat their surfaces as a continuum of classical spins.
Keeping this in mind, we investigate
classical Heisenberg-coupled spins on a deformable
torus section in the presence of topological spin solitons.

The continuum limit of the Heisenberg Hamiltonian for classical
ferromagnets or antiferromagnets for isotropic spin-spin coupling
is the nonlinear $\sigma$-model
\cite{BelavinPolyakov,Trimper,Chakravarty,Haldane,Fradkin}.
The total Hamiltonian for a deformable, magnetoelastically coupled
manifold is given by
$\gtsHm=\gtsHm_{magn}+\gtsHm_{el}+\gtsHm_{m-el}$,
where $\gtsHm_{magn}$, $\gtsHm_{el}$ and $\gtsHm_{m-el}$
represent the magnetic, elastic and magnetoelastic energy,
respectively.
In the present paper we will focus on the magnetic part
and the elastic part only.
For the nonlinear $\sigma$-model, the magnetic energy
on a curved surface $\gtsSurface$, in curvilinear coordinates,
is given by
\cite{VSDTS,TSGF}
\begin{equation}
\label{Hm/mg}
\gtsHm_{magn}=
\gtsJ\!{\iint_\gtsSurface}\!\sqrt{g}\,{\gtsd\Omega}\:
g^{ij}h_{\alpha\beta}\partial_{i}n^{\alpha}\partial_{j}n^{\beta},
\end{equation}
where $\gtsJ$ denotes the coupling energy between neighboring spins.
The order parameter $\widehat{{\mathbf n}}$ is the local magnetization
unit vector specified by a point on the sphere $S^2$. The metric
tensors $(g_{ij})$ and $(h_{\alpha\beta})$ describe respectively
the support surface $\gtsSurface$ and the order parameter manifold:
as customary, ${\gtsd\Omega}$ represents the surface area and
$g$ the determinant $\det(g_{ij})$.

First, let us consider the nonlinear $\sigma$-model on a rigid
torus section. The most natural representation of a torus is given
in cylindrical coordinates $(\rho,\xi,z)$:
\begin{equation}
\rho=R+r\cos \varphi,\quad z=r\sin \varphi,
\end{equation}
where the \emph{rotating radius} $R$ and the \emph{axial radius} $r$
must verify \mbox{$0<r<R$},
while the angle $\varphi$ varies from $-\pi$ to $\pi$.
Nevertheless, for our purposes we will use
a more suitable representation
\cite{Zhongcan}
\begin{equation}
\rho=\frac{a\sinh b}{\cosh b -\cos \eta},
\quad z=\frac{a\sin \eta}{\cosh b -\cos \eta},
\end{equation}
where the new constant parameters $a$ and $b$ are both real and positive,
while the new angle $\eta$ varies from $-\pi$ to $\pi$.
The relations
\begin{equation}
\label{coords/pp/par}
a=\sqrt{\!\left(R+r\right)\!\left(R-r\right)}
\quad\text{and}\quad
\cosh b=\frac{R}{r}
\end{equation}
allow a simple geometrical interpretation for the new parameters:
let us call $a$ the \emph{geometric radius}
and $b$ the \emph{eccentric angle}.
Conversely, the natural parameters $R$ and $r$ satisfy
\begin{equation}
\label{coords/n/par}
R=\frac{a}{\tanh b}
\quad\text{and}\quad
r=\frac{a}{\sinh b}.
\end{equation}
Furthermore, the transformation yields
\begin{subequations}
\begin{eqnarray}
\tan{\scriptstyle{\frac{1}{2}}}\eta
&=&\tanh{\scriptstyle{\frac{1}{2}}}b\;\tan{\scriptstyle{\frac{1}{2}}}\varphi
,\\
&=&\sqrt{\frac{R-r}{R+r}}\tan{\scriptstyle{\frac{1}{2}}}\varphi.
\end{eqnarray}
\end{subequations}
One can easily check that the metric on a torus
in \emph{peri-polar coordinates} $(\xi,\eta)$ is given by
\begin{equation}
g=\frac{a^2}{(\cosh b -\cos \eta)^2}
\left[
\sinh^2 b\ \gtsd\xi\!\otimes\!\gtsd\xi+\gtsd\eta\!\otimes\!\gtsd\eta
\right],
\end{equation}
therefore $g^{\xi\eta}=g^{\eta\xi}=0$ and we have
\begin{equation}
\label{T/g/nested}
g^{\xi\xi}\!\sqrt{g}=\frac{1}{\sinh b},
\quad g^{\eta\eta}\!\sqrt{g}={\sinh b}.
\end{equation}
From now on, we restrict ourselves only to a section of the torus:
the angle $\xi$ about the $z$-axis will vary
from $-\Delta\xi$ to $\Delta\xi$ where \mbox{$0<\Delta\xi<\pi$}.

As usual, the local magnetization $\widehat{{\mathbf n}}$ is described
by its polar coordinates $(\gtsMTheta,\gtsMPhi)$, then the metric
on the Heisenberg sphere is given by
\begin{equation}
h=\gtsd\gtsMTheta\!\otimes\!\gtsd\gtsMTheta
+\sin^2 \gtsMTheta\ \gtsd\gtsMPhi\!\otimes\!\gtsd\gtsMPhi.
\end{equation}

Assuming homogeneous boundary conditions
at both ends \mbox{($\gtsMTheta=0\left[\pi\right]$}
as \mbox{$\xi\to\pm\Delta\xi$)}
allows to map each boundary of the section to a point:
thus we compactify our torus section into the sphere $S^2$.
Consequently, the mapping of our support to the order parameter
manifold is classified by the homotopy group $\Pi_2\left(S^2\right)$
which is isomorphic to $\mathbb{Z}$:
spin configurations may be classified according to
their homotopy class \cite{BelavinPolyakov,Bogomolnyi}.

Henceforth, without loss of generality,
only toroidal symmetric configurations
\mbox{($\partial_{\eta}\gtsMTheta=\partial_{\xi}\gtsMPhi=0$)}
will be considered.
Thus the magnetic Hamiltonian (\ref{Hm/mg}) becomes
\begin{equation}
\label{Hm/mg/xi}
\gtsHm_{magn}=
\gtsJ\!
\int\limits_{-\Delta\xi}^{+\Delta\xi}\!\!\gtsd\xi\!\!
\int\limits_{-\pi}^{+\pi}\!\!\gtsd\eta\,
\left[
\frac{\gtsMTheta_{\xi}^2}{\sinh b}
+\sinh b \sin^2 \gtsMTheta\ \gtsMPhi_{\eta}^2
\right],
\end{equation}
where a subscript stands for differentiation.
Rescaling the angle about $z$-axis
in equation (\ref{Hm/mg/xi}) gives:
\begin{equation}
\label{Hm/mg/zt}
\gtsHm_{magn}=
\gtsJ\!
\int\limits_{-\Delta\zeta}^{+\Delta\zeta}\!\!\gtsd\zeta\!\!
\int\limits_{-\pi}^{+\pi}\!\!\gtsd\eta\,
\left[
\gtsMTheta_{\zeta}^2 +\sin^2 \gtsMTheta\ \gtsMPhi_{\eta}^2
\right],
\end{equation}
where $\zeta\equiv\sinh b\ \xi$ and
$\Delta\zeta\equiv\sinh b\ \Delta\xi$.
The Euler-Lagrange equations
corresponding to (\ref{Hm/mg/zt}) are:
\begin{subequations}
\label{EL/Hm/mg/zt}
\begin{eqnarray}
\gtsMPhi_{\eta\eta}&=&0,\label{EL/Hm/mg/zt/MPhi}\\
\gtsMTheta_{\zeta\zeta}&=&
\gtsMPhi_{\eta}^2 \sin \gtsMTheta \cos \gtsMTheta.
\label{EL/Hm/mg/zt/MTheta}
\end{eqnarray}
\end{subequations}
\begin{subequations}
\label{EL/Hm/mg/zt/nested}
From (\ref{EL/Hm/mg/zt/MPhi}), it follows that
\begin{equation}
\label{EL/Hm/mg/zt/MPhi/nested}
\gtsMPhi_{\eta}=q_{\eta}\qquad q_{\eta}\in\mathbb{Z}.
\end{equation}
Substituting this into (\ref{EL/Hm/mg/zt/MTheta}) and
rescaling again the rotating angle, we get the
sine-Gordon (\textsc{sg}) equation
\begin{equation}
\label{SG/MTheta}
\gtsMTheta_{\varrho\varrho}=\sin \gtsMTheta \cos \gtsMTheta,
\end{equation}
where $\varrho\equiv \gtsq_{\eta}\zeta$,
or $\varrho\equiv \gtsq_{\eta}\sinh b\ \xi$.
\end{subequations}

Equation (\ref{SG/MTheta}) may be integrated once to yield
\begin{equation}
\label{SG/res/int1}
\gtsMTheta_{\varrho}^2=\sin^2 \gtsMTheta +\gtsm
\qquad \gtsm\in\left[0,+\infty\right).
\end{equation}
The limit case $\gtsm=0$ corresponds to
the self-duality equation\cite{TSGF}.
Performing the change of variable
$\sin\gtsMTheta=\JacobiDN\!\left(\gtsu\mid 1+\gtsm\right)$
where $\JacobiDN$ is a Jacobi elliptic function \cite{Abramowitz},
the differential equation becomes $\gtsu_{\varrho}^2=1$.
Let us denote by $\gtsSG\!\left(\cdot\mid\gtsm\right)$
the increasing solution of (\ref{SG/MTheta})
specified by the \emph{parameter} $\gtsm$ and subject to
the boundary condition $\gtsSG\!\left(0\mid\gtsm\right)=\frac{\pi}{2}$.
Readily, we get:
\begin{equation}
\label{SG/res/Jacobi/sin}
\sin\gtsSG\!\left(\varrho\mid\gtsm\right)=
\JacobiDN\!\left(\varrho\mid 1+\gtsm\right).
\end{equation}
Simple calculations lead to the cosine variant:
\begin{subequations}
\label{SG/res/Jacobi/cos}
\begin{eqnarray}
\label{SG/res/Jacobi/cos/crude}
\cos\gtsSG\!\left(\varrho\mid\gtsm\right)
&=&-\sqrt{1\!+\!\gtsm}\:
  \JacobiSN\!\left(\varrho\mid1\!+\!\gtsm\right),\\
\label{SG/res/Jacobi/cos/nested}
&=&-\JacobiSN\!\left(\varrho\sqrt{1\!+\!\gtsm}%
  \mid1/{1\!+\!\gtsm}\right).
\end{eqnarray}
\end{subequations}
Thus, the general \textsc{sg} solution, in natural coordinates, is
\begin{equation}
\label{SG/res/gsol}
\theta\!\left(\varrho\right)=
\gtsSign\gtsSG\!\left(\varrho\mid\gtsm\right)+\gtsConst
\qquad\gtsSign=\pm{1}.
\end{equation}
Further the function $\gtsSG\!\left(\cdot\mid\gtsm\right)$ satisfies
\begin{equation}
\label{SG/rel/qp}
\gtsSG\!\left(\varrho+2\gtsq\gtsK_{\gtsSG}\mid\gtsm\right)=
\gtsSG\!\left(\varrho\mid\gtsm\right)+\gtsq\pi
\qquad\gtsq\in\mathbb{Z},
\end{equation}
where the \emph{quasi quarter-period} $\gtsK_{\gtsSG}$
is related to the complete elliptic integral of
the first kind $\EllipticK$ by
\cite{Abramowitz}
\begin{subequations}
\label{SG/qp/ei}
\begin{eqnarray}
\label{SG/qp/ei/crude}
\gtsK_{\gtsSG}\!\left(\gtsm\right)
&=&\EllipticK\!\left(1\!+\!\gtsm\right),\\
\label{SG/qp/ei/nested}
&=&\frac{1}{\sqrt{1\!+\!\gtsm}}\:
  \EllipticK\!\!\left(\frac{1}{1\!+\!\gtsm}\right).
\end{eqnarray}
\end{subequations}

Using the solutions of equation (\ref{SG/MTheta}),
the \mbox{$\gtsq_{\xi}\pi$}-soliton configuration consistent with the
boundary conditions can be obtained easily.
Up to an irrelevant additive multiple of $\pi$, we have
\begin{subequations}
\label{SConf}
\begin{equation}
\label{SConf/MTheta}
\gtsMTheta\!\left(\xi\right)=
\gtsSign\gtsSG\!\left(\gtsq_{\eta}\sinh b\ \xi\mid\gtsm\right)
-\delta_{\text{even},\gtsq_{\xi}}{\textstyle{\frac{\pi}{2}}},
\end{equation}
where the parameter $\gtsm$ is given by
\begin{equation}
\label{SConf/m}
\gtsm=\gtsK_{\gtsSG}^{-1}\!\!\left(
\frac{\gtsq_{\eta}}{\gtsq_{\xi}}\Delta\xi\:\sinh b
\right).
\end{equation}
\end{subequations}

The magnetic energy $\gtsE_{magn}$ of the above configuration (\ref{SConf})
may be compared with the corresponding topological minimum energy $\gtsTE$
which does not depend on the geometry of the support manifold.
Performing the Bogomol'nyi's decomposition\cite{Bogomolnyi} yields
\begin{equation}
\label{Bgml/E}
\gtsTE=8\pi\gtsJ\vert\gtsQ\vert,
\end{equation}
where the topological charge (i.e. the winding number)
$\gtsQ$ equals to $\gtsq_{\xi}\gtsq_{\eta}$.
Let $\gtsEr_{magn}$ denote the ratio $\gtsE_{magn}/\gtsTE$.
A straightforward calculation shows that $\gtsEr_{magn}$ depends
on the parameter $\gtsm$ only; we have
\begin{subequations}
\label{Bgml/Er}
\begin{eqnarray}
\gtsEr_{magn}
\label{Bgml/Er/crude}
&=&\EllipticE\!\left(1+\gtsm\right)+
  {\textstyle\frac{1}{2}}\gtsm\:
  \EllipticK\!\left(1+\gtsm\right),\\
\label{Bgml/Er/nested}
&=&\sqrt{1+\gtsm}\left[
  \EllipticE\!\!\left(\frac{1}{1\!+\!\gtsm}\right)\!-\,
  {\textstyle\!\frac{1}{2}}\frac{\gtsm}{1\!+\!\gtsm}\:
  \EllipticK\!\!\left(\frac{1}{1\!+\!\gtsm}\right)
\right],
\end{eqnarray}
\end{subequations}
which increases strictly from $1$ to $\infty$
with respect to $\gtsm$;
$\EllipticE$ is the complete elliptic integral of
the second kind\cite{Abramowitz}.
To avoid any unnecessary complication, we will consider the
quantity \mbox{${\gtsq_{\eta}\Delta\xi}/\!{\gtsq_{\xi}}$} fixed.
Therewith, according to (\ref{SConf/m}), $\gtsEr_{magn}$ is
a decreasing function of the eccentric angle $b$.
Clearly, the minimum energy $\gtsTE$ is only reached
when $b$ tends to infinity,
i.e., when our support becomes an infinite rigid cylinder.
This is in agreement with results obtained for the rigid cylinder
\cite{VSDTS,TSGF}.
In other words, the non-satisfaction of the Bogomol'nyi's inequality
is due to the geometry of the support manifold,
hence the expression: geometrical frustration.
Therefore, here, the geometric frustration is induced by introducing
a second non-vanishing local curvature.
Furthermore, notice that only the eccentric angle,
which measures the balance between
the rotating radius $R$ and the axial radius $r$,
does tune the spread of the soliton.

Next, let us relax the rigidity constraint and consider
the nontrivial spin configuration on an elastic torus section.
Accordingly, the soliton will try to minimize its
magnetic energy $\gtsE_{magn}$ to the minimum energy $\gtsTE$
by deforming the elastic support\cite{VSDTS,TSGF,CIGFMS}.

Since the eccentric angle $b$ appears as the relevant
geometric parameter and the rotating angle $\xi$ as the relevant
curvilinear coordinate, we relax $b$ with respect to $\xi$ and
write
\begin{equation}
\label{aT/ea}
b\!\left(\xi\right)=b_{0}+\gtsTL\!\left(\xi\right),
\end{equation}
where $b_{0}$ represents the \emph{spontaneous eccentric angle}
and the function $\gtsTL$ describes local deformations.
The metric on the deformable manifold in this case
remains orthogonal, and (\ref{T/g/nested}) reads
\begin{equation}
\label{aT/g/nested}
g^{\xi\xi}\!\sqrt{g}\!=\!\!\frac{1}{\sqrt{\,\sinh^2 b+\gtsTL_{\xi}^2}},
\quad g^{\eta\eta}\!\sqrt{g}\!=\!\!{\sqrt{\,\sinh^2 b+\gtsTL_{\xi}^2}}.
\end{equation}

As the problem is quasi-one-dimensional,
the magnetoelastic energy $\gtsHm_{m-el}$ merely renormalizes
the spin coupling energy $\gtsJ$ in the magnetic energy
$\gtsHm_{magn}$\cite{CHC}.
Therefore, we add to the nonlinear
$\sigma$-model Hamiltonian (\ref{Hm/mg})
only the elastic energy which is essentially
stored in the bending of the deformable support
\cite{Canham,HelfrichZN1,MAPeterson,SMMS}:
\begin{equation}
\label{Hm/el}
\gtsHm_{el}={\textstyle{\frac{1}{2}}}\gtsKc\!
{\iint_\gtsSurface}\!\sqrt{g}\,{\gtsd\Omega}\:
\left(\gtsMC-\gtsMC_{0}\right)^2.
\end{equation}
Here the constant $\gtsKc$ denotes the \emph{bending rigidity},
$\gtsMC$ represents the mean curvature \cite{Struik},
and the \emph{spontaneous mean curvature} $\gtsMC_{0}$ tends
to bias the mean curvature for recovering the spontaneous shape.
Assuming small and smooth deformations and expanding to second order
in $\gtsTL$, $\gtsTL_{\xi}$ and $\gtsTL_{\xi\xi}$
lead to \cite{HST}
\begin{subequations}
\begin{equation}
\label{aT/Hm/el}
\gtsHm_{el}=
{\textstyle{\frac{1}{2}}}\pi\gtsKc\gtsCst_{2}\!
\int\limits_{-\Delta\zeta}^{+\Delta\zeta}\!\!\!\gtsd\zeta\,\gtsTL^2,
\end{equation}
where
\begin{equation}
\gtsCst_{2}=
\frac{1+\sinh b_{0}\cosh b_{0}\left(3+\sinh^2 b_{0}\right)}%
  {\sinh^4 b_{0}}.
\end{equation}
\end{subequations}

Before deriving the Euler-Lagrange equation for the total
Hamiltonian $\gtsHm=\gtsHm_{magn}+\gtsHm_{el}$,
we calculate the magnetic energy associated with
the nontrivial spin configuration (\ref{SConf}).
Expanding to second order in $\gtsTL$ and $\gtsTL_{\xi}$ the
relations (\ref{aT/g/nested}) enables to rewrite (\ref{Hm/mg/zt})
as follows
\begin{eqnarray}
\label{aT/Hm/mg}
\gtsHm_{magn}&&=
\gtsJ\!
\int\limits_{-\Delta\zeta}^{+\Delta\zeta}\!\!\gtsd\zeta\!\!
\int\limits_{-\pi}^{+\pi}\!\!\gtsd\eta\,
\left[
\vphantom{\frac{\gtsTL^2\gtsMTheta_{\zeta}^2}{\sinh^2b_{0}}}
\left(1+{\textstyle\frac{1}{2}}\gtsTL^2\right)
\left[\gtsMTheta_{\zeta}^2 +\sin^2 \gtsMTheta\ \gtsMPhi_{\eta}^2
\right]\right.\nonumber\\
-&&\left(\coth b_{0}\,\gtsTL+{\textstyle\frac{1}{2}}\gtsTL_{\zeta}^2\right)
\left[\gtsMTheta_{\zeta}^2 -\sin^2 \gtsMTheta\ \gtsMPhi_{\eta}^2
\right]\nonumber\\
&&\left.\hphantom{\gtsJ\!\int\!\!\gtsd\zeta\int\!\!\gtsd}
+\frac{\gtsTL^2\gtsMTheta_{\zeta}^2}{\sinh^2 b_{0}}
\right].
\end{eqnarray}
On the other hand,
according to (\ref{SG/res/int1}) and (\ref{SG/res/Jacobi/sin}),
the spin configuration (\ref{SConf}) verifies
\begin{eqnarray*}
\gtsMTheta_{\zeta}^2&-&\sin^2\gtsMTheta\ \gtsMPhi_{\eta}^2=
\gtsm\gtsq_{\eta}^2,\\
\gtsMTheta_{\zeta}^2&=&
\left(1+\gtsm\right)\gtsq_{\eta}^2
\left[1-\JacobiSN^2\!\!\left(\gtsq_{\eta}\zeta\mid 1+\gtsm\right)
\right].
\end{eqnarray*}
Inserting these relations into (\ref{aT/Hm/mg}) and simplifying,
we obtain
\begin{eqnarray}
\label{SConf/Hm/mg/2nd}
\gtsHm_{magn}&&=
2\pi\gtsJ\gtsq_{\eta}^2\!
\int\limits_{-\Delta\zeta}^{+\Delta\zeta}\!\!\gtsd\zeta\,
\left[
\vphantom{\frac{2}{\sinh^2 b_{0}}\frac{1+\gtsm}{2+\gtsm}}
\left(2+\gtsm\right)-\gtsm
\left[\coth b_{0}\,\gtsTL+{\textstyle\frac{1}{2}}\gtsTL_{\zeta}^2
\right]\right.\nonumber\\
-&&\left(1+\gtsm\right)
\left[2+\coth^2 b_{0}\,\gtsTL^2\right]
\JacobiSN^2\!\!\left(\gtsq_{\eta}\zeta\mid 1+\gtsm\right)\nonumber\\
&&\left.\hphantom{2\pi\gtsJ\gtsq}
+{\textstyle\frac{1}{2}}
\left(2+\gtsm+2\frac{1+\gtsm}{\sinh^2 b_{0}}
\right)\gtsTL^2
\right].
\end{eqnarray}
The Euler-Lagrange equation for the problem takes the following form
\begin{subequations}
\label{EL/TL}
\begin{equation}
\label{EL/TL/eqn}
\gtsTL_{\varrho\varrho}+
\left[\left(1+\gtsLm\right)\!\gtsLA
-\gtsLm\gtsLB\;\JacobiSN^2\!\!\left(\varrho\mid\gtsLm\right)
\right]\gtsTL
=\sqrt{\gtsLm}\ \gtsLj,
\end{equation}
where we have set
\begin{eqnarray}
\gtsLm&=&1+\gtsm,\\
\gtsLA&=&\frac{1}{\gtsm\gtsq_{\eta}^2}
\left[1+\frac{2}{2+\gtsm}
\left(
\frac{1+\gtsm}{\sinh^2 b_{0}}
+\frac{\gtsCst_{2}}{\gtsJr\gtsq_{\eta}^2}
\right)\right],\\
\gtsLB&=&2\frac{\coth^2 b_{0}}{\gtsm\gtsq_{\eta}^2},\\
\gtsLj&=&\frac{\coth b_{0}}{\gtsq_{\eta}^2\sqrt{1+\gtsm}}.
\end{eqnarray}
\end{subequations}
Here we have introduced the \emph{relative coupling energy}
$\gtsJr\equiv\gtsJ/\gtsKc$ and the variable
$\varrho\equiv\gtsq_{\eta}\zeta$ (see (\ref{SG/MTheta})).
The linear inhomogeneous second-order differential equation
(\ref{EL/TL/eqn}) is related to the well-known homogeneous
(Jacobian) Lam\'e's equation which occurs in several physical
contexts \cite{HST,Arscott}.
We have found no direct treatment of
(\ref{EL/TL/eqn}) in the literature.
However, an approach based on the derivation of
the Lam\'e functions \cite{HST,Arscott}
allows to find a particular solution denoted by
$\gtsIL\!\left(\cdot\mid\gtsLm;\gtsLA,\gtsLB,\gtsLj\right)$.
Therefore, if small and smooth deformations are assumed, a
suitable deformation function $\gtsTL$ is given by
\begin{equation}
\label{aT/TL}
\gtsTL\!\left(\xi\right)=
\gtsIL\!\left(
\gtsq_{\eta}\sinh b_{0}\ \xi\mid 1+\gtsm;\gtsLA,\gtsLB,\gtsLj
\right).
\end{equation}

The geometric frustration becomes evident in considering
the Euler-Lagrange equation (\ref{EL/TL/eqn}):
$\gtsTL=0$ (i.e. the rigid torus section) is not a solution.
Now let us turn our attention to the deformation mechanism.
As we have seen, the magnetic energy $\gtsE_{magn}$ of
the configuration (\ref{SConf}) is a decreasing function
of the eccentric angle $b$:
therefore, for sufficiently small and smooth deformations,
the soliton tries to increase the eccentric angle $b$.
From a physical point of view and according to (\ref{coords/pp/par}),
the soliton tends to collapse the torus section
($R\to a^{+}$ and $r\to0^{+}$ as $b\to\infty$).
On the other hand,
the elastic Hamiltonian (\ref{Hm/el}) tries to maintain
the spontaneous shape of the deformable support:
the bending rigidity tends to ``attract'' the eccentric angle $b$
to the spontaneous eccentric angle $b_{0}$.
The underlying physics makes a good sense:
rigidity prevents the support from a collapse \ldots
To sum up, any alteration of the eccentric angle $b$
increases the elastic energy whereas
any increasing (resp. decreasing) of the eccentric angle $b$
decreases (resp. increases) the magnetic energy
---at least for small and smooth deformations.
Consequently,
the competition between the magnetic energy and the elastic energy
induces an augmentation of the eccentric angle $b$.
Further more,
according to (\ref{Hm/mg/zt}) the soliton energy
is essentially localized in the spread zone:
the alteration is less important in the spin flip region.

\begin{figure}
\begin{center}
\centerline{\epsfxsize=\linewidth\epsfbox{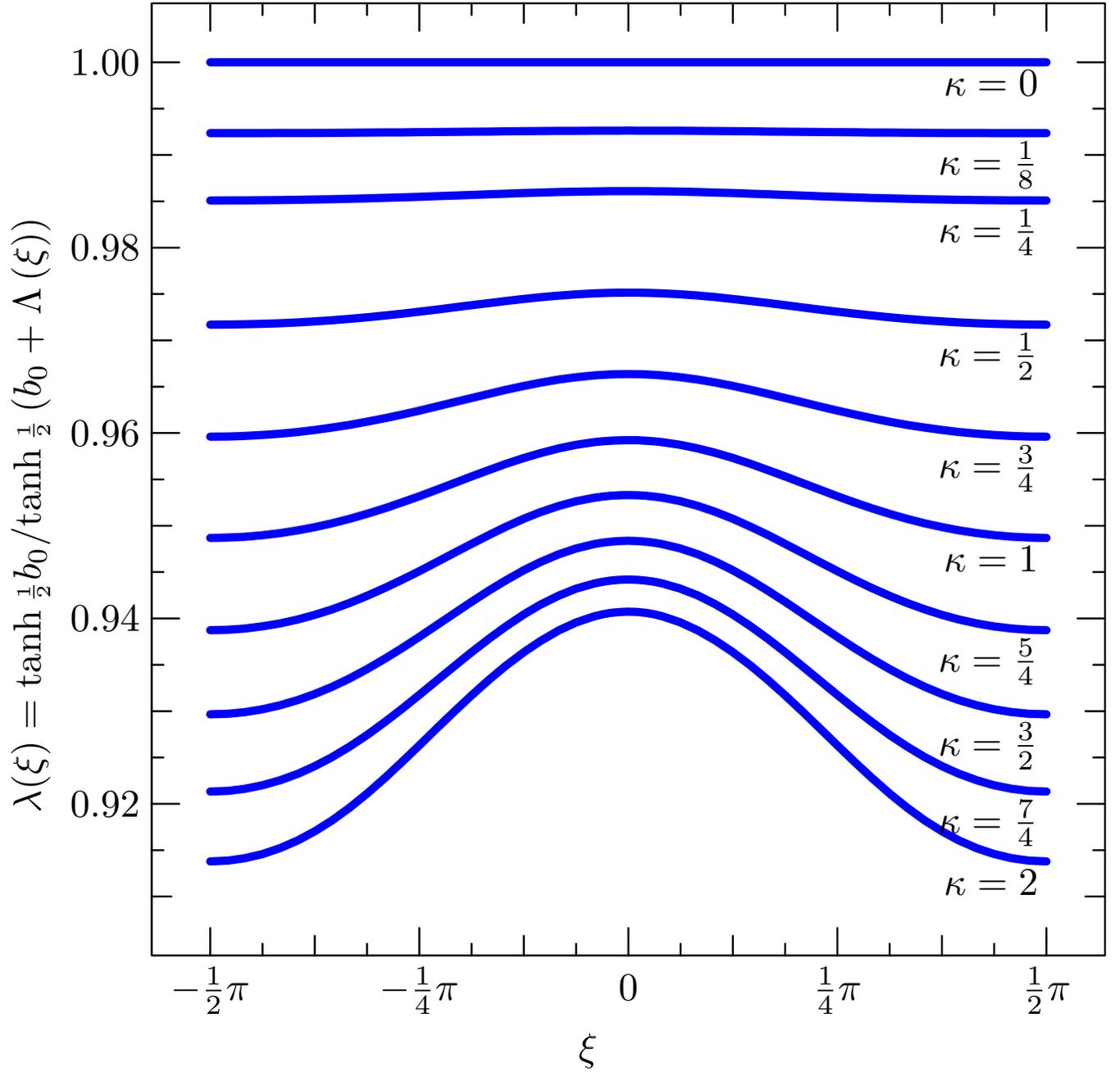}}
\end{center}
\caption[relative dilatation $\gtsrd\!\left(\xi\right)$]{%
  The relative dilatation $\gtsrd\!\left(\xi\right)$ as defined
  in (\ref{FR/rd/def}) corresponding to the deformation
  function (\ref{aT/TL}) associated with
  the $\pi$-Clifford torus section
  ($2\Delta\xi=\pi$ and $\sinh b_{0}=1$)
  in presence of a $\pi$-soliton (\ref{SConf})
  versus the rotating angle $\xi$ for different values of
  the relative coupling energy $\gtsJr\equiv\gtsJ/\gtsKc$:
  the case $\gtsJr=0$ corresponds to the rigid torus section.}
\label{fig/RDPC/k}
\end{figure}

To understand the geometric meaning of the frustration release,
let us introduce the \emph{outer radius} $\gtsOR=R+r$
and the \emph{inner radius} $\gtsIR=R-r$.
Their geometrical interpretation is transparent.
Now, noting that $\gtsOR\,\gtsIR=a^2$,
we define the \emph{relative dilatation} $\gtsrd$ by
\begin{equation}
\label{FR/rd/def}
\gtsrd=\frac{\gtsOR}{\gtsOR_{0}}=%
  \left(\frac{\gtsIR}{\gtsIR_{0}}\right)^{-1},
\end{equation}
where $\gtsOR_{0}$ and $\gtsIR_{0}$ are related to
the spontaneous shape.
The transformation relationships (\ref{coords/n/par}) allows
to write a readable relation between
the relative dilatation $\gtsrd$ and the eccentric angle $b$:
\begin{equation}
\gtsrd=
\frac{\tanh{\scriptstyle{\frac{1}{2}}}b_{0}}
  {\tanh{\scriptstyle{\frac{1}{2}}}b}.
\end{equation}
Clearly, the augmentation of the eccentric angle $b$ leads to
a global shrinking whereas a local swelling arises
where the spins twist, as shown in Fig.~\ref{fig/RDPC/k}.

In conclusion, we have shown that the Euler-Lagrange equation
for the nonlinear $\sigma$-model on a torus section
with homogeneous boundary conditions is the sine-Gordon equation:
the spread of the topological soliton configurations is tuned by
the torus eccentricity.
Besides, the model presents a geometrical frustration induced by
the torus eccentricity.
As announced, the frustration release is shown to lead
to a novel geometric effect:
the torus section is \textit{globally shrunk}
and a swelling appears in the region of the soliton.
The full torus problem will be considered
in a subsequent paper\cite{HST}.

We acknowledge fruitful discussions with J. Mourad.


\end{multicols}

\end{document}